\documentclass[twocolumn,showpacs]{revtex4}
\usepackage{graphicx}
\usepackage{dcolumn}
\usepackage{amsmath}
\begin{document}

\author{Matthew L. Wallace and B\'ela Jo\'{o}s}
\email{bjoos@uottawa.ca}
\affiliation{Ottawa-Carleton Institute for Physics, University of
Ottawa Campus,\\ Ottawa, Ontario, Canada K1N 6N5\\}

\date{\today}

\title{Shear-induced overaging in a polymer glass}

\begin{abstract}
\vspace*{0.2cm} A phenomenon recently coined as ``overaging''
implies a slowdown in the collective (slow) relaxation modes of a
glass when a transient shear strain is imposed. We are able to
reproduce this behavior in simulations of a supercooled polymer melt
by imposing instantaneous shear deformations. The increases in
relaxation times $\Delta \tau_{1/2}$ rise rapidly with deformation,
becoming exponential in the plastic regime.  This ``overaging'' is
distinct from standard aging. We find increases in pressure,
bond-orientational order and in the average energy of the inherent
structures ($<e_{IS}>$) of the system, all dependent on the size of
the deformation. The observed change in behavior from elastic to
plastic deformation suggests a link to the physics of the ``jammed
state''.
\end{abstract}
 \maketitle

In recent years, there has been considerable progress geared towards
understanding how glasses respond to shear. Phenomena such as shear
thinning and ``rejuvenation'' are common when shear flow is imposed.
Unlike crystals, glasses ``age,'' meaning that their state depends
on their history \cite{Kob97,Kob00,Kob00b}. When a glass falls out
of equilibrium, it evolves over very long timescales. It has been
found that relaxation in supercooled liquids often depends on
cooperative and spatially correlated motion, meaning the dynamics
are heterogeneous \cite{Gebremichael01}. Much research has also been
devoted to developing a wide-reaching, coherent theory which can
explain the jammed state \cite{Liu98,Trappe01}. It has been found
that such a state is characterized by the appearance of a yield
shear stress and can be achieved by changing the load, density or
temperature of a system. In addition, it has been found that the
concept of random close packing is somewhat ill-defined and in the
so-called jammed state, one can increase the degree packing at the
expense of randomness, or vice-versa, thus allowing for a variety of
possible jammed configurations \cite{Torquato00,Kansal02}.

The concept of jammed state also provides insight as to the effect
of stress on a glass. When imposing a continuous shear strain, one
can rejuvenate the glass, effectively wiping out the memory of the
system \cite{Sollich97,Barrat03}. In this sense, the glassy state is
particularly sensitive to shearing. In 2002, Viasnoff and Lequeux
published experimental results for colloidal suspensions, showing
that after imposing transient shear, one obtains simultaneous
overaging and rejuvenation, since the relaxation times are altered
in a non-trivial manner \cite{Viasnoff02}. In this case, overaging
means that relaxation times of the system become longer more quickly
than is normally the case. Their results are qualitatively explained
through Bouchaud's Trap Model and the related Soft Glassy Rheology
(SGR) model \cite{Bouchaud92,Sollich97}. In essence, this
phenomenological approach relies on the distribution of relaxation
times $\tau$ (or equivalently, of potential wells of different
depths) which can be over- or underpopulated when applying a
transient shear or a temperature step. Using a similar perspective,
Lacks and Osborne have argued, in a zero temperature study, that the
so-called ``overaging'' is different in nature from ordinary aging,
in that the minima visited can be similar in both cases, but not
identical \cite{Lacks04}. It has also become clear that it is
difficult to induce decreases in energy by strain in well-annealed
glasses, since the disappearance of energy minima does not usually
lead the system to a lower potential well
\cite{Lacks01,Lacks04,Utz00}.

In this letter, we report on some of the characteristics of this
overaging from a more general perspective. We find that overaging is
present and easily observed in simulations of a common glass-forming
polymer model simply by imposing relatively small, instantaneous
shear deformations. After a certain waiting time $t_w$, one can
observe an unambiguous slowing-down in the decorrelation of particle
positions, just as was seen experimentally \cite{Viasnoff02}. We use
the term ``overaging'' to refer only to the longer relaxation times,
not to any decrease in energy of the system (as in the case of
ordinary aging). We have identified two distinct regimes of
overaging, corresponding to elastic and plastic deformations. There
is a rapid increase in relaxation times
$\Delta\tau_{1/2}(\epsilon)$, which contrasts with the associated
increase in the average energy of the internal structure of the
system ($<e_{IS}>$), as introduced by Stillinger and Weber
\cite{Stillinger82}. We concur with the recent study, finding that
overaging is quite distinct from aging in its regular sense
\cite{Lacks04}. In our case, we show that it is possible to obtain
longer relaxation times without the system's slow evolution towards
more energetically favorable configurations. In addition, the yield
shear strain of the material plays a key role in this phenomenon.
Finally, we resolve an increase in order in the system, primarily
associated with the elastic energy stored in the system.

The simulations are performed by molecular dynamics, using the
velocity Verlet algorithm, combined with velocity-rescaling to
achieve constant temperature \cite{Allen}. We adopt the bead-spring
model to simulate the polymer melt \cite{Kremer90}, each chain
consisting of 10 monomers linked together through a
finitely-extensible nonlinear elastic (FENE) potential of the form
${U}_{FENE} (r_{ij}) = 0.5 k R_0^2 \log \left[ 1 - \left( r_{ij} /
R_0 \right)^2 \right]$, where $k=30 \varepsilon _{LJ}/\sigma ^2$ and
$R_0=1.5\sigma$. In addition, all particles interact via a truncated
and shifted Lennard-Jones (LJ) potential, $U_{LJ} \left( {r_{ij} }
\right) = 4\varepsilon _{LJ}\left[ {\left( {\sigma / r_{ij} }
\right)^{12} - \left( {\sigma / r_{ij} } \right)^6} \right]$ with
the cutoff radius set at $2.5\sigma$. For simplicity, all data are
presented in reduced LJ units based on the mass $m$, the radius of
each particle $\sigma$ and the LJ energy scale $\varepsilon_{LJ}$.
Each MD step corresponds to $0.005$ reduced time units. The samples
have 105 chains, so a total of $N=1050$ particles,  and periodic
boundary conditions (PBCs) are applied in all directions. We use up
to 40 independent samples in order to average the results. The
combination of the FENE and LJ potentials causes two competing
length scales, thus inhibiting crystallization and producing a
typical fragile glass \cite{Wallace04,Binder03,Baschnagel00}. In a
previous paper, we have identified a glass transition (GT) based on
the simulation timescales used, as well as a rigidity transition
(RT) located just below the GT \cite{Wallace04}. We also observed
that the shear modulus $\mu$, computed via instantaneous
deformations, is highly dependent on the size of the shear
deformation $\epsilon$. Specifically, larger values of $\epsilon$
can be characterized as irreversible or plastic deformations which
alter the structure (and energy landscape) of the system. In such
cases, the residual stress is reduced leading to smaller $\mu$
values. In contrast to the recent work of Lacks and Osborne
\cite{Lacks04}, we have a well-annealed system, as well as a
non-zero temperature. This means that the system has the ability to
explore a \textit{range} of energy wells (as opposed to a single
one) through thermal activation.

\begin{figure}
\resizebox{3.25in}{2.5in}{\includegraphics{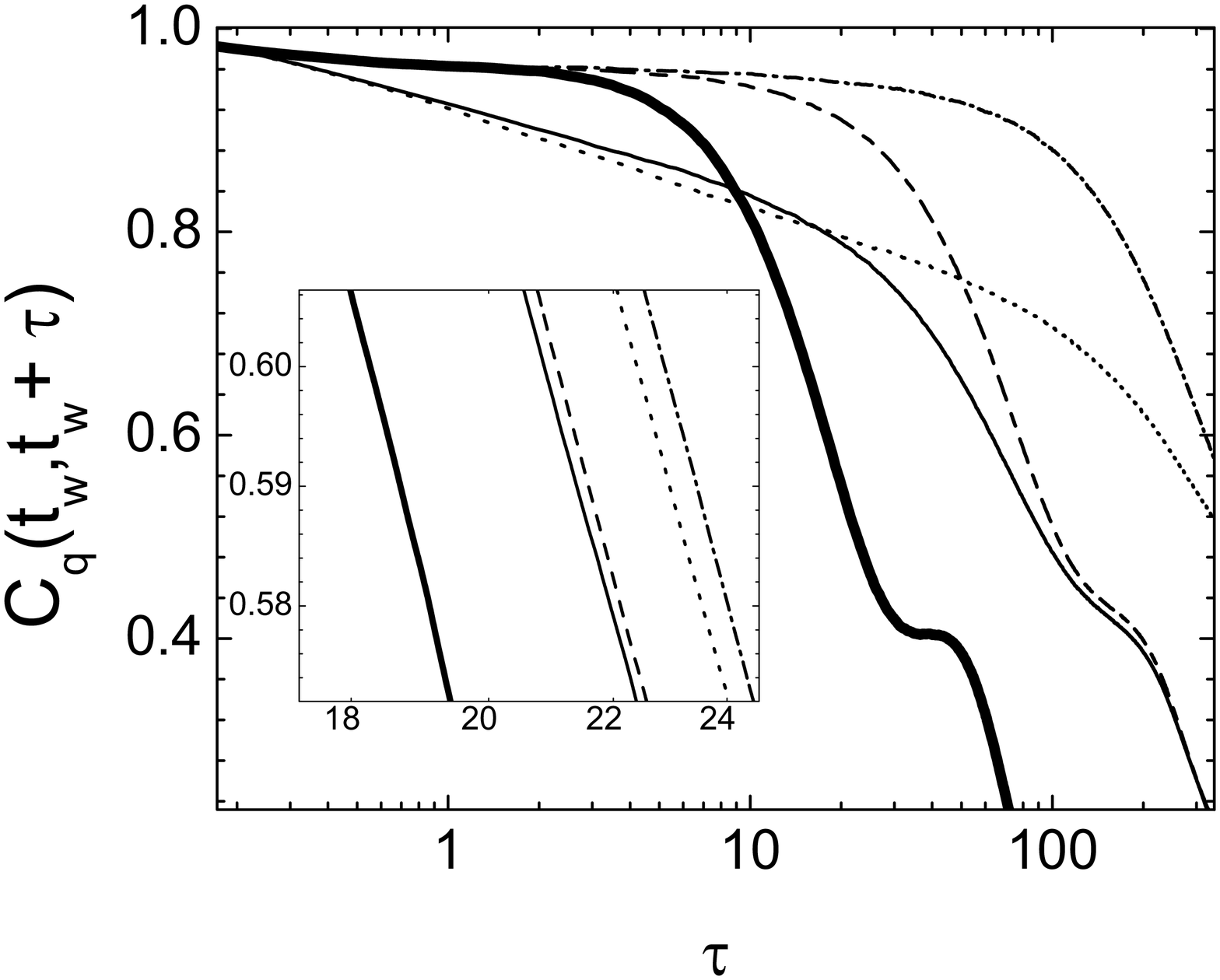}}
\caption{Correlation function of Eq. (\ref{incoherent}) showing the
effects of repeated shear, and different waiting times after each
one, on the slow relaxation modes of the system, compared with the
reference (unsheared) case (bold line). Legend for both the inset
and main frame: first deformation at $t_w=0$ (solid line) and
$t_w=10^3$ (dashed); second deformation at $t_w=0$ (dotted) and
$t_w=10^3$ (dash-dot). Main frame: large shear deformations
($\epsilon=0.2$). Note the initial combination of rejuvenation and
overaging, followed by pure overaging. Inset: Small shear
($\epsilon=0.05$). The relaxation changes, although in a very
different manner, and causes small overaging.}\label{corrfig}
\end{figure}

All samples are set at $T=0.44$ and are initially at identical
pressures in the neighborhood of 0.57 in the rigid phase (near the
onset of rigidity which occurs below the GT \cite{Wallace04}). They
are created via a slow compression of the simulation box, followed
by a damped-force algorithm to realize the NPT ensemble
\cite{Allen}. By definition, these systems are out of equilibrium,
but we start with samples that have been allowed to evolve
considerably and see no evidence of aging on the timescales of our
computer experiments. Starting with a cube of side $L$, an affine
shear deformation $\varepsilon _{xy}$ is applied in the $x$
direction, in a plane with its normal along the $y$ direction.  An
atom initially in position $(x,y,z)$ is displaced to $(x+ \epsilon
_{xy}y, y, z)$. The boundaries of the simulation box consequently
are shifted  for $x_{min}$ from 0 to $\epsilon _{xy} y $ , and for
$x_{max}$ from $L$ to $\epsilon _{xy} y  + L$. For a given system,
the shearing is repeated in the five other directions, substituting
$xy$ by $yz$,$-xy$, etc. Individual samples lack symmetry and
therefore the various deformations will not usually give the same
stress components. Once the system has been allowed to relax for a
fairly long time $t_w$, this process can be repeated on the
previously deformed sample, either deforming it further or returning
it to its original shape. Both methods yield identical increases in
relaxation times.

In order to monitor the relaxation in the system, we compute the
two-time, ``self'' part of the incoherent scattering function,

\begin{equation}\label{incoherent}
C_{\bf q}(t_w+\tau,t_w)=\frac{1}{N}\sum\limits_{j = 1}^N \exp[i {\bf
q}{\cdot}({\bf r}_j(t_w+\tau)-{\bf r}_j(t_w))]
\end{equation}

\noindent where the wavevector {\bf q} is close to the first peak in
the structure factor $S({\bf q})$, $t_w$ is the time elapsed since
the deformation. This ``two-time'' correlation function has been
proven useful in investigating aging, because of the different
timescales in relaxation \cite{Kob97,Kob00}. We use $\tau_{1/2}$,
the time it takes for $C_{\bf q}$ to decrease to 0.5, in order to
gauge the local decorrelation in the system. The effect of the
deformations is seen in Fig. \ref{corrfig}. As was found in Ref.
\cite{Viasnoff02}, for large deformations ($\epsilon \gtrsim 0.1$),
there is an initial combination of overaging and rejuvenation, as
relaxation times corresponding to both high and low-energy states
are overpopulated. Eventually, the initial shape of the function is
recovered since only the slow relaxations remain overpopulated.
Interestingly, one can repeat this process a number of times,
achieving the same results after every applied deformation. Our
focus is not on the transient effects, but rather on the behavior
after a relatively long $t_w$. As seen in the inset of Fig.
\ref{corrfig}, small shear also causes overaging but is the result
of a simpler physical mechanism.

\begin{figure}
\resizebox{3.25in}{2.5in}{\includegraphics{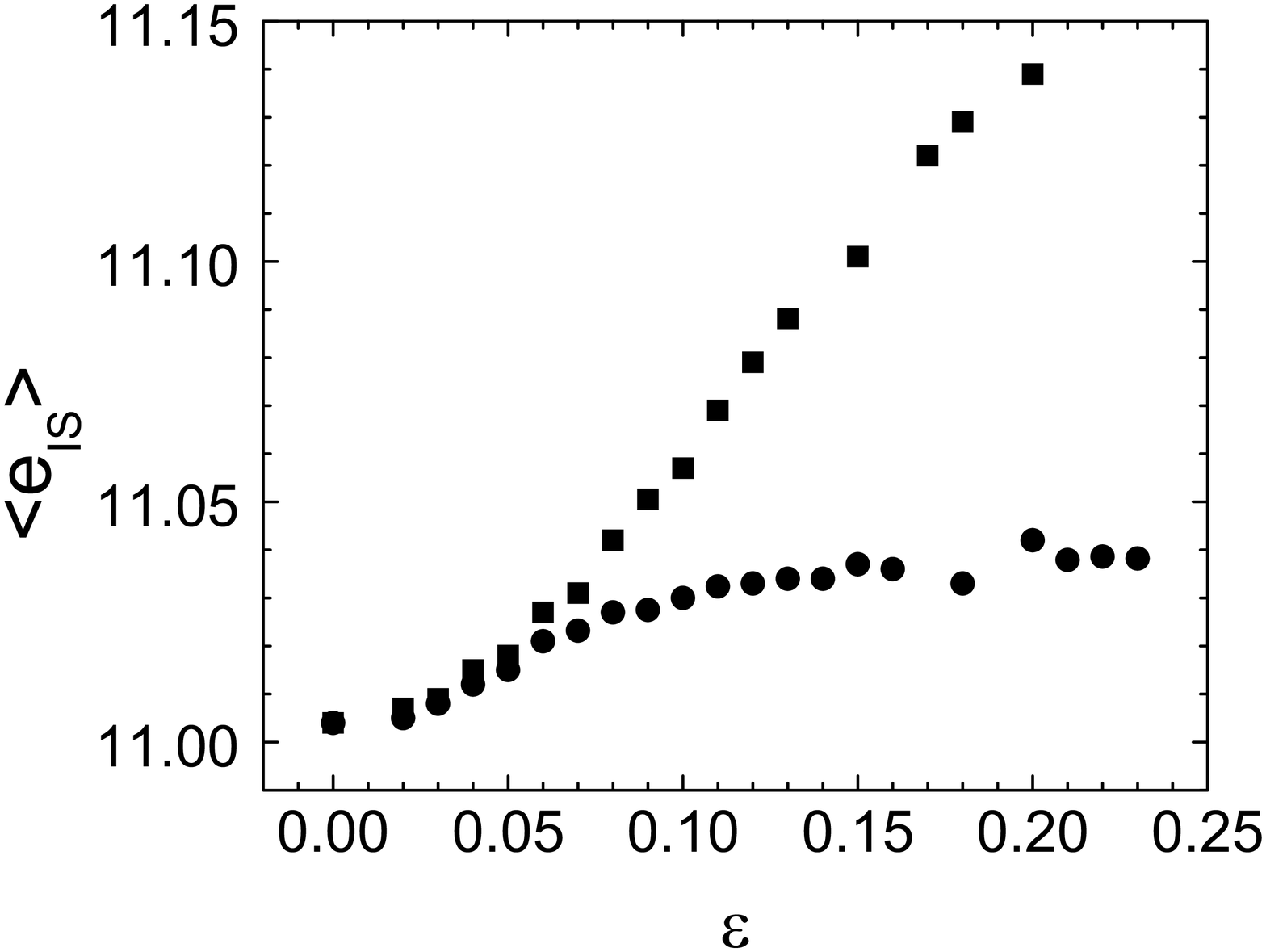}}
\caption{Energy of the inherent structure (IS) of the system,
calculated for various $\epsilon$ immediately after (squares) and
$t_w = 10^3$ (circles) after the deformation. Note that, for large
(plastic) deformations, the inherent structure of the system is able
to evolve.}\label{energyfig}
\end{figure}

\begin{figure}
\resizebox{3.25in}{2.5in}{\includegraphics{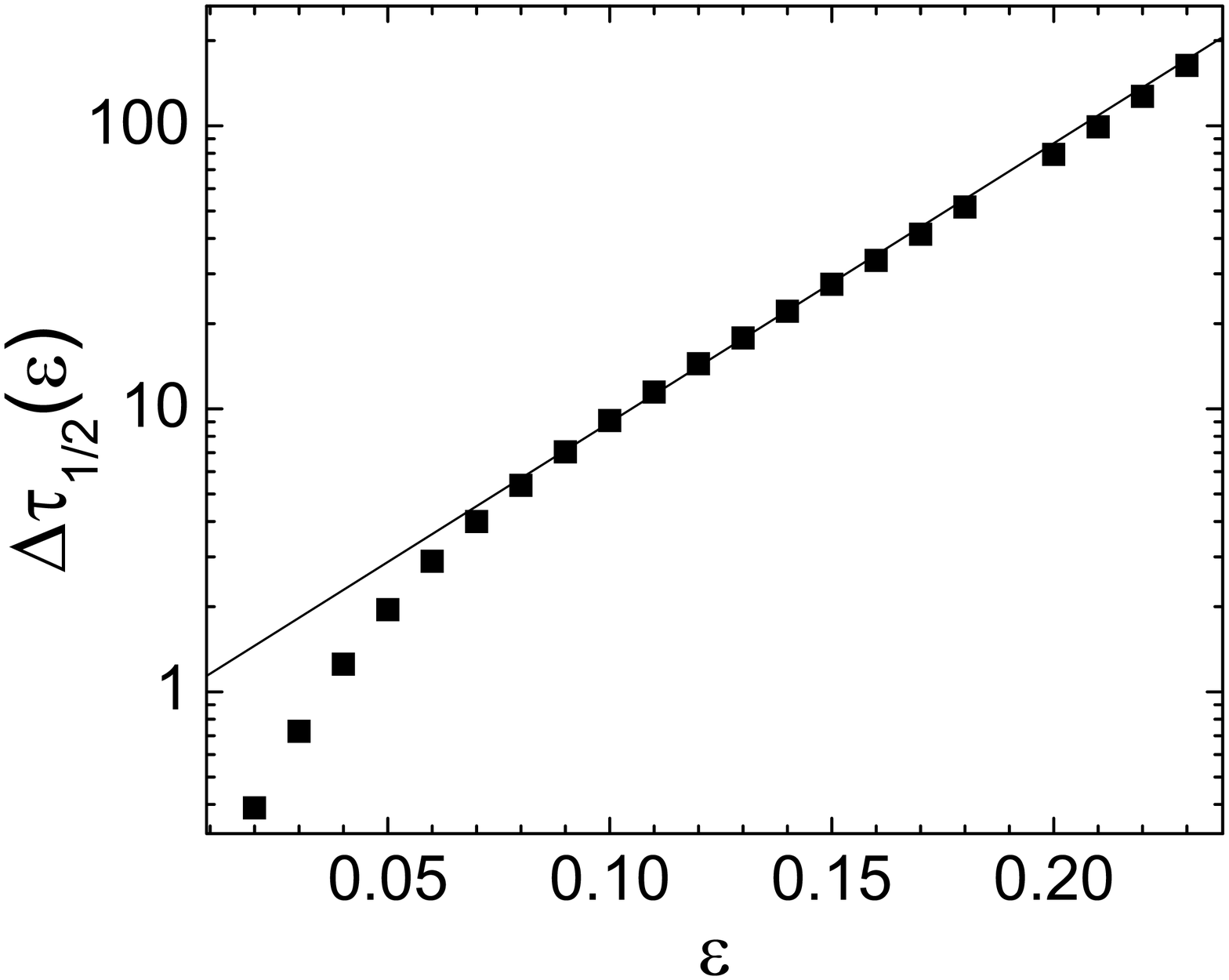}}
\caption{Incremental relaxation times after $t_w=10^3$ for various
$\epsilon$. The line is a guide to the eye and shows an exponential
dependence on $\epsilon$ for plastic deformations.}\label{taufig}
\end{figure}

We start from a well-relaxed system. The strain increases the energy
and appears to raise the energy barriers between available energy
wells, as evidenced by the increased decorrelation times. For small
shear as seen in Fig. \ref{energyfig} there is no major
configurational change. The difference between the $<e_{IS}>$
immediately after and a relatively long time after a given
deformation is minimal. This difference increases rapidly as we
enter the plastic regime where a large part of the strain energy is
relaxed by the configurational changes.  Plastic shear maps the
system onto a very different energy well, from which the system can
escape fairly easily. This is the source of the initial combination
of overaging and rejuvenation observed in Fig. \ref{corrfig} and in
Ref. \cite{Viasnoff02}. After a reasonable waiting time ($t_w =
10^3$), there is a net increase in $\tau _{1/2}$. The characteristic
incremental relaxation time $\Delta\tau_{1/2}(\epsilon)$ is shown
for various shear deformations $\epsilon$ in Fig. \ref{taufig},
after subtracting the reference relaxation time
$\tau_{1/2}(\epsilon=0)$, which is a constant on our simulation
timescales. In the plastic regime, there is a clear exponential
behavior.

To get some inkling of what may be causing the overaging we look at
the structural changes occurring. When applying a relatively large
deformation, the chains are initially stretched from their
equilibrium lengths. But after letting the system relax for
approximately $10^3$ time units, the radial distribution function
and the average radius of gyration of the chains return to their
``undeformed'' values. A more sensitive measure of structural change
is the local bond-orientational order parameter
$Q_{6,\mathrm{local}}$ which uses higher-order spherical harmonics
and is defined as \cite{Steinhardt83,Torquato00,Kansal02}

\begin{equation}\label{q6}
Q_{6,\mathrm{local}}\equiv \sum\limits_{j = 1}^N
\left(\frac{4\pi}{13} \sum\limits_{m = -6}^6 \left | \frac{1}{n_b}
\sum\limits_{i = 1}^{n_b^j} Y_{6m}(\theta_i , \phi_i ) \right |^2
\right )^{1/2}
\end{equation}

\begin{figure}
\resizebox{3.25in}{2.5in}{\includegraphics{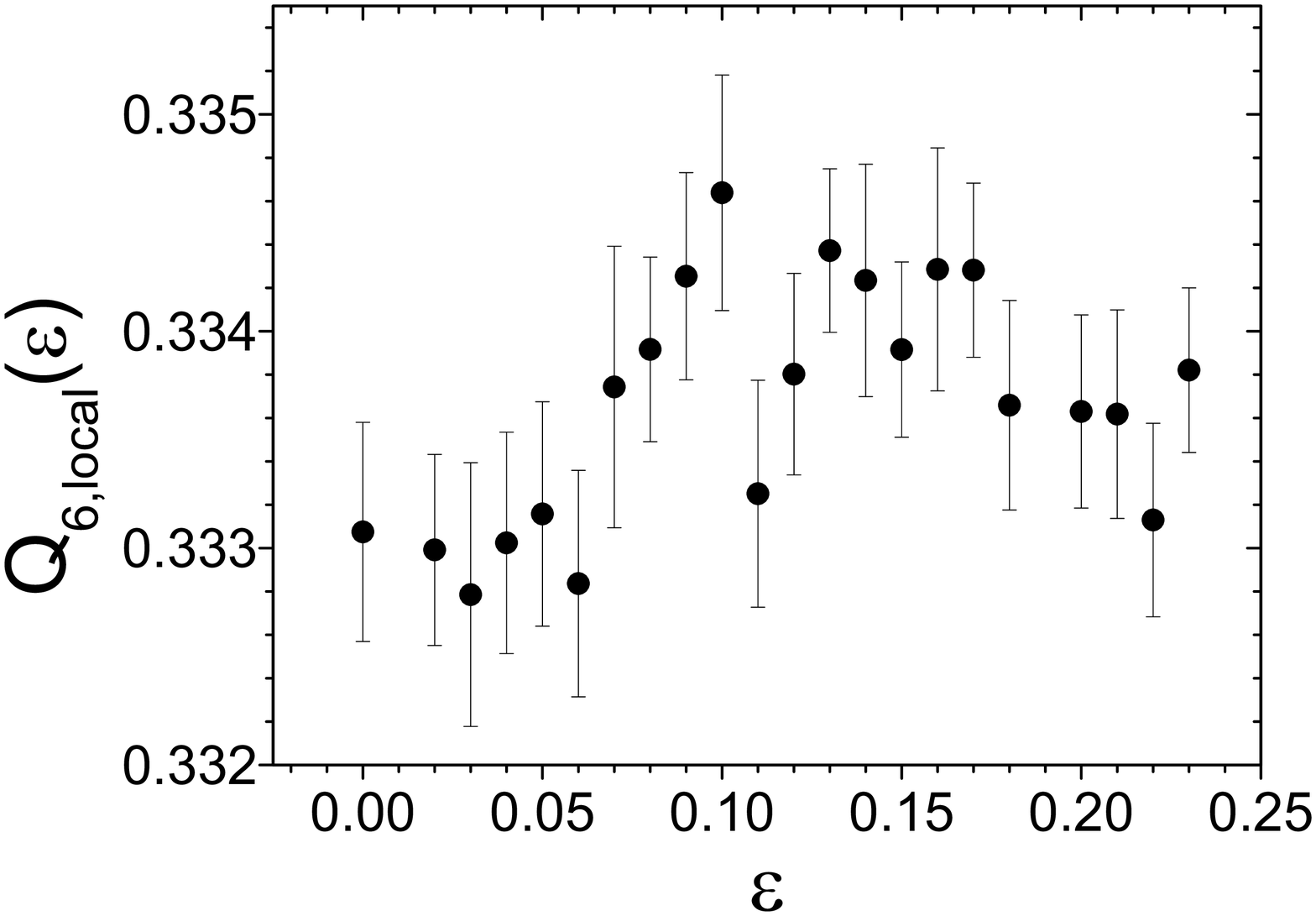}}
\caption{Evolution of the local bond-orientational order parameter
$Q_{6,\mathrm{local}}$ at $t_w = 10^3$ after the deformation. While
the statistics are poor, there is a clear increase in
$Q_{6,\mathrm{local}}$, at least for $\epsilon \lesssim 0.1$. Data
based on 40 samples deformed twice, and treating each deformation as
independent.}\label{orderfig}
\end{figure}

\begin{figure}
\resizebox{3.25in}{2.5in}{\includegraphics{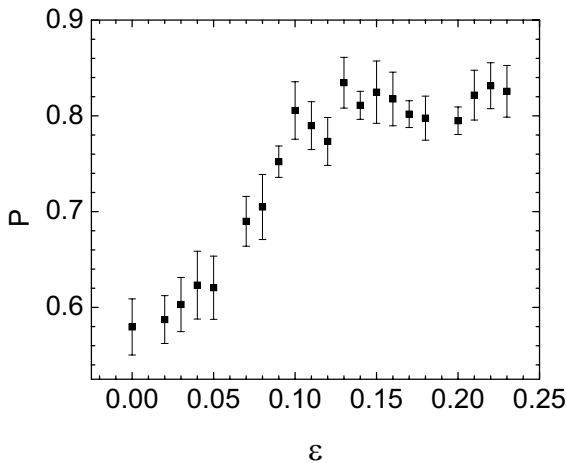}}
\caption{Pressure in the system, calculated for various $\epsilon$
at $t_w = 10^3$ after the deformation. The pressure only increases
for small (reversible) deformations.} \label{presfig}
\end{figure}

\noindent Indeed, $Q_{6,\mathrm{local}}$ increases slightly as we
increase $\epsilon$, while $\epsilon\lesssim 0.1$, followed by a
plateau or slight decrease (see Fig. \ref{orderfig}), noting that
subsequent deformations do not have a large effect on the order
parameter.  This ordering, is most apparent for small shear, indicating
that it is directly related to the elastic energy stored in the system.
Increased order is usually associated with a lower
energy state.  In our polymeric system, however, the potential
energy is dominated by the stiff intra-chain bonds. The increased
order is at the expense of the potential energy, and leads to a
pressure increase  (see Fig. \ref{presfig}) (also observed in LJ
binary mixtures \cite{Utz00}). The increased $\tau _{1/2}$ is an
indication that there are larger energy barriers between
configurations, i.e. that the strain  ``jams'' the system; and in the case
of plastic shear, reorganizes the potential wells. The dynamics are
also affected. Once again, the two regimes of shear strain are very
distinct. Small strains have little or no effect, while large
strains reduce the ``heterogeneity'' in the system (\emph{i.e.};
there are less deviations from a Gaussian distribution of particle
displacements \cite{Gebremichael01}) and more particles become
``mobile'' \cite{Wallace1}. This, in turn, causes a decrease in the
shear modulus $\mu$ of the system, \emph{i.e.} shear softening
\cite{Wallace04}. Increased mobility is not incompatible with
increased relaxation time. Average waiting times are dominated by
the contributions from the largest $\tau _{1/2}$ of the
``metabasins'' whereas the mobility is controlled only by the most
mobile particles \cite{Wallace1,Heuer00}.

An interesting aspect of this study is the similarity with
``tapping'' experiments on granular matter \cite{Nowak97} and glassy
systems \cite{Fierro02}, which can produce jammed structures. Common
features in the structure and force distributions of granular and
``glassy'' jamming have also recently become apparent
\cite{Silbert02}. Just as tapping granular matter can both increase
or decrease the density, transient or instantaneous shear can
produce a variety of effects on the glassy system, while continuous
shear always ``unjams'' or rejuvenates it. In conclusion, this is
not aging in the conventional sense (the system has not aged more
quickly); $<e_{IS}>$ does not decrease and neither does the
mobility. The very similar behavior observed in colloids
\cite{Sollich97,Viasnoff02}, the present polymer glass
\cite{Wallace04}, and binary Lennard Jones mixtures \cite{Utz00}
suggests a universal behavior and unique origin of the overaging,
observed so far in two of these systems.  Shear produces changes in
the microstructure which lead to increased relaxation times. This
may be put to profitable use in some applications. It appears also
to be a way to increase ordering. Finally, repeated applications of
the deformations yield increasingly larger relaxation times, and may
generate a glassy state of practical interest, whose properties are
yet to be fully understood.

We acknowledge the support of the Natural Sciences and Engineering
Council (Canada) and stimulating discussions with Dan Vernon and
Michael Plischke.


\begin{thebibliography} {99}

\bibitem{Kob00b} W. Kob, F. Sciortino and P. Tartaglia, Europhys.
Lett. {\bf 49}, 590 (2000).
\bibitem{Kob97} W. Kob and J.-L. Barrat, Phys. Rev. Lett. {\bf
78}, 4581 (1997).
\bibitem{Kob00} W. Kob and J.-L. Barrat, Eur.
Phys. J. B {\bf 13}, 319 (2000).
\bibitem{Gebremichael01} W. Kob., C. Donati, S.J. Plimpton, P.H. Poole
and S.C. Glotzer, Phys. Rev. Lett. {\bf 79}, 2827 (1997); Y.
Gebremichael, T.B. Schr{\o}der, F.W. Starr and S.C. Glotzer, Phys.
Rev. E {\bf 64}, 051503 (2001); K. Vollmayr-Lee, W. Kob, K. Binder
and A. Zippelius, J. Chem. Phys. {\bf 116}, 5158 (2002).
\bibitem{Liu98} A.J. Liu and S.R. Nagel, Nature {\bf 396},
21 (1998).
\bibitem{Trappe01} V. Trappe, V. Prasad, L. Cipelletti, P.N. Segre
and D.A. Weitz, Nature {\bf 411}, 772 (2001).
\bibitem{Torquato00} S. Torquato, T.M. Truskett and P.G.
Debenedetti, Phys. Rev. Lett. {\bf 84}, 2064 (2000).
\bibitem{Kansal02} A.R. Kansal, S. Torquato and F.H. Stillinger,
Phys. Rev. E {\bf 66}, 041109 (2002).
\bibitem{Barrat03} J.-L. Barrat, J. Phys.: Cond. Mat. {\bf 15} S1
(2003).
\bibitem{Sollich97} P. Sollich, F. Lequeux, P. H\'{e}braud and M.E.
Cates, Phys. Rev. Lett. {\bf 78}, 2020 (1997).
\bibitem{Viasnoff02} V. Viasnoff and F. Lequeux, Phys. Rev. Lett. {\bf 89},
065701 (2002); V. Viasnoff, S. Jurine and F. Lequeux, Faraday
Discussions {\bf 123}, 253 (2003).
\bibitem{Bouchaud92} J.-P. Bouchaud, J. Physique {\bf 2}, 1705
(1992).
\bibitem{Lacks01} D.J. Lacks, Phys. Rev. Lett {\bf 87}, 225502 (2001).
\bibitem{Lacks04} D.J. Lacks and M.J. Osborne, Phys. Rev. Lett. {\bf 93},
255501 (2004).

\bibitem{Stillinger82} F.H. Stillinger and T.A. Weber, Phys. Rev.
A {\bf 25}, 978 (1982).
\bibitem{Kremer90} K. Kremer and G.S. Grest, J. Chem. Phys. {\bf 92}, 5057
(1990).
\bibitem{Wallace04} M. L. Wallace, B. Jo\'{o}s and M. Plischke,
Phys. Rev. E {\bf 70}, 041501 (2004).
\bibitem{Baschnagel00} J. Baschnagel, C. Bennemann,
W. Paul, and K. Binder, J. Phys.: Condens. Matter {\bf 12}, 6365
(2000).
\bibitem{Binder03} K. Binder, J. Baschnagel, and W. Paul, Prog.
Polym. Sci. 28, {\bf 115} (2003).
\bibitem{Allen} M.P. Allen and D.J. Tildesley, \textit{Computer Simulations
of Liquids} (Oxford University Press, New York, 1987).
\bibitem{Utz00} M. Utz, P.G. Debenedetti and F.H. Stillinger,
Phys. Rev. Lett {\bf 84}, 1471 (2000).
\bibitem{Steinhardt83} P.J. Steinhardt, D.R. Nelson, and M.
Ronchetti, Phys. Rev. B {\bf 28}, 784 (1983).
\bibitem{Thorpe00} M.F. Thorpe, D.J. Jacobs, M.V. Chubynsky and
J.C. Phillips, J. Non-Cryst. Solids {\bf 266-269}, 859 (2000); A.
Huerta and G.G. Naumis, Phys. Rev. Lett. {\bf 90}, 145701 (2003).
\bibitem{Wallace1} M.L. Wallace and B. Jo\'{o}s, to be published.
\bibitem{Heuer00} S. B\"{u}chner and A. Heuer, Phys. Rev. Lett.
{\bf84}, 2168 (2000); B. Doliwa and A. Heuer, Phys. Rev. E {\bf 67}
030501(R) (2003).
\bibitem{Nowak97} E.R. Nowak, J.B. Knight, M.L. Povinelli, H.M.
Jaeger and S.R. Nagel, Powder Technol. {\bf 94}, 79 (1997).
\bibitem{Fierro02} A. Fierro, M. Nicodemi and A. Coniglio,
Europhys. Lett. {\bf 59}, 642 (2002).
\bibitem{Silbert02} L.E. Silbert, D. Ertas, G.S. Grest, T.C.
Halsey and D. Levine, Phys. Rev. E {\bf 65}, 051307 (2002).

\end{thebibliography}
\end{document}